\documentclass[10pt,letterpaper,twocolumn,english]{article}
\usepackage[T1]{fontenc}
\usepackage[latin9]{inputenc}
\usepackage{babel}
\usepackage{amsmath}
\usepackage{amssymb}
\usepackage{graphicx}
\usepackage[unicode=true,pdfusetitle,
 bookmarks=true,bookmarksnumbered=false,bookmarksopen=false,
 breaklinks=false,pdfborder={0 0 1},backref=section,colorlinks=false]
 {hyperref}
\hypersetup{
 draft}
\usepackage{breakurl}

\makeatletter

\usepackage{ol2}

\makeatother

\begin{document}
\twocolumn[ 
\title{Radial mode dependence of optical beam shifts}

\author{N. Hermosa,$^{1,*}$ Andrea Aiello,$^{2,3}$ and J.P. Woerdman$^1$}

\address{
$^1$Huygens Laboratory, Leiden University, P.O. Box 9504, 2300 RA Leiden, The Netherlands
\\
$^2$Max Planck Institute for the Science of Light, Günther-Scharowsky-Strasse 1/Bau 34, 91058 Erlangen, Germany\\
$^3$Institute of Optics, Information and Photonics, University Erlangen-Nürnberg, Staudtstrasse 7/B2, 91058 Erlangen, Germany\\
$^*$Corresponding author: hermosa@physics.leidenuniv.nl
}

\begin{abstract}It is known that orbital angular momentum (OAM) couples the Goos-Hänchen and Imbert-Fedorov shifts. Here, we present the first study of these shifts when the OAM-endowed $LG_{\ell,p}$ beams have higher-order radial mode index (\emph{p>0}).  We show theoretically and experimentally that the angular shifts are enhanced by \emph{p} while the positional shifts are not. \end{abstract}


 ] 

\noindent A bounded beam of light may experience a displacement and/or
a deflection with respect to the geometric optics prediction upon
reflection \cite{AielloOLRole2008,LiUnified2006}. These shifts are
known as the Goos-Hänchen (GH)\cite{GoosHanchen1947} and the Imbert-Fedorov
(IF)\cite{Fedorov1955,Imbert1972} shifts. A shift in the plane of
incidence indicates that the beam underwent a GH shift while a shift
that is perpendicular to the plane of incidence connotes an IF shift.
Moreover, these shifts are further distinguished because of their
positional and angular parts\cite{AielloOLRole2008}. The positional
shift is the displacement of the center point of reflection while
an angular shift happens when the beam deflects and propagates with
a slight deviation from the law of reflection. The angular shift is
responsible for the increasing beam excursion from the geometric optics
prediction with propagation. All these shifts are in the order 0.1-10
wavelengths for the positional shifts and micro- to milliradians for
the angular shifts.

These shifts may occur simultaneously or separately. Their presence
depends on the polarization of the incident beam\cite{AielloOLRole2008,LiUnified2006,Imbert1972,GoosHanchen1947,Fedorov1955},
the index gradient seen by the beam\cite{MeranoOpEx2007,HermosaOL2011,MenardPRB2010,BermanPRE2002},
the beam\textquoteright{}s divergence (opening angle of the beam)\cite{AielloOLRole2008,AielloPRA2009,MeranoNatPhotonics2009}
and, as emphasized recently, on the modal structure of the beam\cite{BliokhOL2009,MeranoPRA2010,AilleoOl2011,DasguptaOptComm2006,GollaPramana2011}.
In works that involve modal structures, the Laguerre-Gaussian beam
is used as the definitive beam because of its orbital angular momentum
(OAM). The OAM of light couples the four beam shifts (the positional
GH and IF shifts and their angular counterparts). First described
in \cite{BliokhOL2009}, we subsequently proposed a matrix which was
validated with experiments, that neatly summarizes this mixing\cite{MeranoPRA2010}.
But since we considered only the case of a LG radial mode index that
is zero $\left(p=0\right)$, that matrix is only explicitly dependent
on the \textit{azimuthal} mode index, $\ell$. The most important
result of \cite{MeranoPRA2010} is the measurement of the OAM-induced
positional GH and IF and the OAM-affected angular GH and IF shifts
for $\ell=-1,1$. We extended the measurement for $|\ell|>1$ in \cite{HermosaSPIE2011}.
The shifts are found to be linearly enhanced by the OAM.

In this Letter, we answer the question to what degree the complete
spatial mode of a beam affects these shifts. We first do theoretical
calculations and extend our previous matrix formulation to the case
of $\mathit{p>0}$. Next we report experiments on shifts associated
with higher-order mode Laguerre-Gaussian modes of different azimuthal
and radial mode indices. 

Laguerre-Gaussian modes are a set of solutions of the paraxial wave
equation of the form given by

\begin{align}
u_{p}^{\ell}\left(r,\phi\right) & =\left[\frac{2p!}{\pi\left(p+|\ell|\right)}\right]^{1/2}\exp\left(i\ell\phi\right)\exp\left(\frac{i}{2}\frac{r^{2}}{z-iL}\right)\nonumber \\
 & \quad\times\exp\left[-i\left(2p+|\ell|+1\right)\arctan\left(\frac{z}{L}\right)\right]\nonumber \\
 & \quad\times\frac{1}{w\left(z\right)}L_{p}^{\ell}\left(\frac{2r^{2}}{w^{2}\left(z\right)}\right)\left(\frac{\sqrt{2}r}{w\left(z\right)}\right)^{|\ell|}
\end{align}
where $x=r\cos\phi$, $y=r\sin\phi,$ $w_{0}$ is the fundamental
beam waist, $L=k_{0}w_{0}^{2}/2$ is the Rayleigh length, $w\left(z\right)=w_{0}\sqrt{1+\left(z/L\right)^{2}}$
is the beam waist at $z$, $k_{0}$ is the center wave vector, and
$L_{p}^{\ell}(x)$ denotes the generalized Laguerre polynomial of
order $\left(\ell,p\right)$ with $p\geq0$ and $\ell$ as integers.
This beam is characterized by the indices $|\ell|$ and $p$ , where
$\ell$ is the azimuthal mode index and $p+1$ is the number of radial
nodes \cite{Siegman}. There has been little interest in the index
$p$ ; mainly for use in optical manipulation and trapping\cite{KennedyPRA2002,MatsumotoJOSAA2008,ChavezPRL2003,TemperePRA2011}.
However, higher LG modes ($p,|\ell|>0$) are increasingly being noticed,
for instance in gravitational wave detection. As a matter of fact,
some tabletop experiments are being done to test their suitability\cite{ChelkowskiPRD2009,FuldaPRD2010,GranataPRL2010}.

The theoretical derivation of the effect of the radial mode index
follows from \cite{AielloOLRole2008,MeranoPRA2010}. The incident
and reflected beams are assumed to be Gaussian. The incident beam
is then decomposed into plane wave components and the Fresnel reflection
coefficients are applied to the \textit{s} and \textit{p} polarization
components of the wave, respectively. The shift is obtained by summing
all the reflected plane wave components, and taking the centroid of
the intensity distribution.

Following the notation in \cite{MeranoPRA2010} and noting that we
are using a median detector \cite{HermosaQDOL2011,centroid}, we arrive
at a matrix which is explicitly dependent on both $\ell$ and $p$
, given by,

\begin{align}
\left[\begin{array}{c}
\Delta_{GH}^{\ell}\\
\Theta_{IF}^{\ell}\\
\Delta_{IF}^{\ell}\\
\Theta_{GH}^{\ell}
\end{array}\right]= & \left[\begin{array}{cccc}
1 & -2\ell & 0 & 0\\
0 & \xi\left(\ell,p\right) & 0 & 0\\
0 & 0 & 1 & 2\ell\\
0 & 0 & 0 & \xi\left(\ell,p\right)
\end{array}\right] & \left[\begin{array}{c}
\Delta_{GH}^{0}\\
\Theta_{IF}^{0}\\
\Delta_{IF}^{0}\\
\Theta_{GH}^{0}
\end{array}\right]
\end{align}
where $\Delta_{GH}^{0}$ and $\Delta_{IF}^{0}$ are the dimensionless
positional GH and IF shifts, respectively, and $\Theta_{GH}^{0}$
and $\Theta_{IF}^{0}$ are the dimensionless angular GH and IF shifts,
respectively, for a $TEM_{00}$ beam. These are given by,
\begin{align}
\Delta_{GH}^{0}=w_{p}\mathrm{Im}\left(\frac{\partial\ln r_{p}}{\partial\theta}\right)+w_{s}\mathrm{Im}\left(\frac{\partial\ln r_{s}}{\partial\theta}\right) & ,
\end{align}

\begin{align}
-\Theta_{GH}^{0}=w_{p}\mathrm{Re}\left(\frac{\partial\ln r_{p}}{\partial\theta}\right)+w_{s}\mathrm{Re}\left(\frac{\partial\ln r_{s}}{\partial\theta}\right) & ,
\end{align}

\begin{align}
\Delta_{IF}^{0} & =-\frac{a_{p}a_{s}\cot\theta}{R_{p}^{2}a_{p}^{2}+R_{s}^{2}a_{s}^{2}}\times\nonumber \\
 & \quad\left[\left(R_{p}^{2}+R_{s}^{2}\right)\sin\eta+2R_{p}R_{s}\sin\left(\eta-\varphi_{p}+\varphi_{s}\right)\right],
\end{align}

\begin{align}
\Theta_{IF}^{0}=\frac{a_{p}a_{s}\cot\theta}{R_{p}^{2}a_{p}^{2}+R_{s}^{2}a_{s}^{2}}\left[\left(R_{p}^{2}-R_{s}^{2}\right)\cos\eta\right]
\end{align}
with $w_{s/p}=R_{s/p}^{2}a_{s/p}^{2}/\left(R_{p}^{2}a_{p}^{2}+R_{s}^{2}a_{s}^{2}\right)$,
and $r_{s/p}=R_{s/p}\exp\left(i\varphi_{s/p}\right)$ the Fresnel
reflection coefficient evaluated at the incident angle $\theta_{i}$
and $a_{s/p}$ the electric field components. The shifts $x$ and
$y$ are the sum of two contributions and are given by $k_{0}x=\Delta_{GH}^{\ell}+(z/L)\Theta_{GH}^{\ell}$
and $k_{0}y=\Delta_{IF}^{\ell}+(z/L)\Theta_{IF}^{\ell}$, respectively.

Finally, the matrix element $\xi\left(\ell,p\right)$ calculated for
a median detector is $\xi\left(\ell,p\right)=2I_{+}/I_{-}$ where

\begin{align}
I_{\pm}\left(\ell,p\right) & =\sum_{a=0}^{p}\sum_{b=0}^{p}\frac{\left(-1\right)^{a+b}}{a!b!}\left(\begin{array}{c}
p+|\ell|\\
p-a
\end{array}\right)\left(\begin{array}{c}
p+|\ell|\\
p-b
\end{array}\right)\nonumber \\
 & \quad\times\Gamma\left(a+b+|\ell|+1\pm\frac{1}{2}\right).
\end{align}

It follows from (2) that the positional shift is not affected by the
$p$ index. Only the angular shift components are influenced by it.
Moreover, the matrix elements $M_{22}$ and $M_{44}$ are the same.
This means that the enhancements of the angular GH and IF shifts are
the same.

\begin{figure}[htb]
 \centerline{\includegraphics[width=7cm]{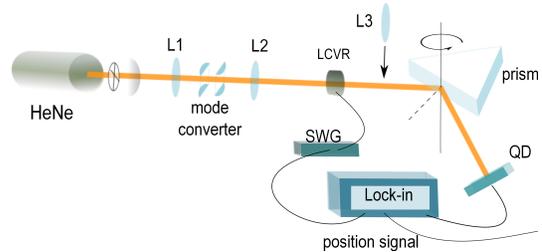}} \caption{(color online) The experimental set-up used to measure the polarization-differential
beam shifts as a function of the incident angle. LCVR is the variable
retarder. SWG is the square wave generator. QD is the quadrant detector.
L1, L2, and L3 are lenses. More details are in the text.}
\end{figure}

Our experimental setup is shown in Fig. 1. We use an open cavity HeNe
laser ($\lambda=632.8nm$) and placed two crossed copper wires ($d\sim90\mu m$)
inside the cavity. The wires force the cavity to emit a higher-order
Hermite-Gaussian mode ($HG_{m,n}$). The integer mode indices $m$
and $n$ can be dialed by properly shifting the intracavity wires.
With an astigmatic mode converter, the $HG_{m,n}$ modes are then
converted into $LG_{\ell,p}$ modes with transformations $\ell=m-n$
and $p=min(m,n)$ as described in \cite{BeijersbergenOptComm1993,AllenPRA1992}.
L1 is the mode matching lens while lens L2 collimates the beam. The
beam then passes through an LC variable retarder (LVCR) which sets
the beam's polarization state. The LCVR (Meadowlark) is driven by
a square wave generator at 2.5 Hz. This wave generator also provides
the reference wave of the lock-in amplifier (EG\&G 5210). The positional
shifts are measured with a collimated beam while the angular shifts
are measured with a beam focused by lens L3 (\textit{f=70 mm}).
The displacement due to the angular shifts increases with the opening
angle of the beam $\theta_{0}$, as $\theta_{0}^{2}/2$ where $\theta_{0}=2/\left(k_{0}\omega_{0}\right)$\cite{MeranoNatPhotonics2009,MeranoOL352010}.

The polarization differential shifts are measured between $45^{0}/-45^{0}$
polarization states. These states are specifically chosen because:
1) the dimensionless angular IF shift $\Theta_{IF}^{0}$ that is responsible
for the $\Delta_{GH}^{\ell}$ and $\Theta_{IF}^{\ell}$ is maximum
at these states; and 2) the contribution of $\Delta_{GH}^{0}$, $\Delta_{IF}^{0}$
, and $\Theta_{GH}^{0}$ either have values equivalent to zero or
cancel to zero upon measurement of the differential shift. These enable
us to isolate the matrix elements $M_{12}$ and $M_{22}$ in (2).

After the polarization conditioning, the beam is made to reflect at
different incident angles $\theta_{i}$ on a BK7 prism ($n=1.51$)
mounted on a $\theta-2\theta$ rotation stage. The position of the
reflected beam is measured by a quadrant detector (QD) whose signal
is fed to the lock-in amplifier. These are done for different combinations
of $\ell$ and $p$.

\begin{figure}[htb]
 \centerline{\includegraphics[width=6.25cm]{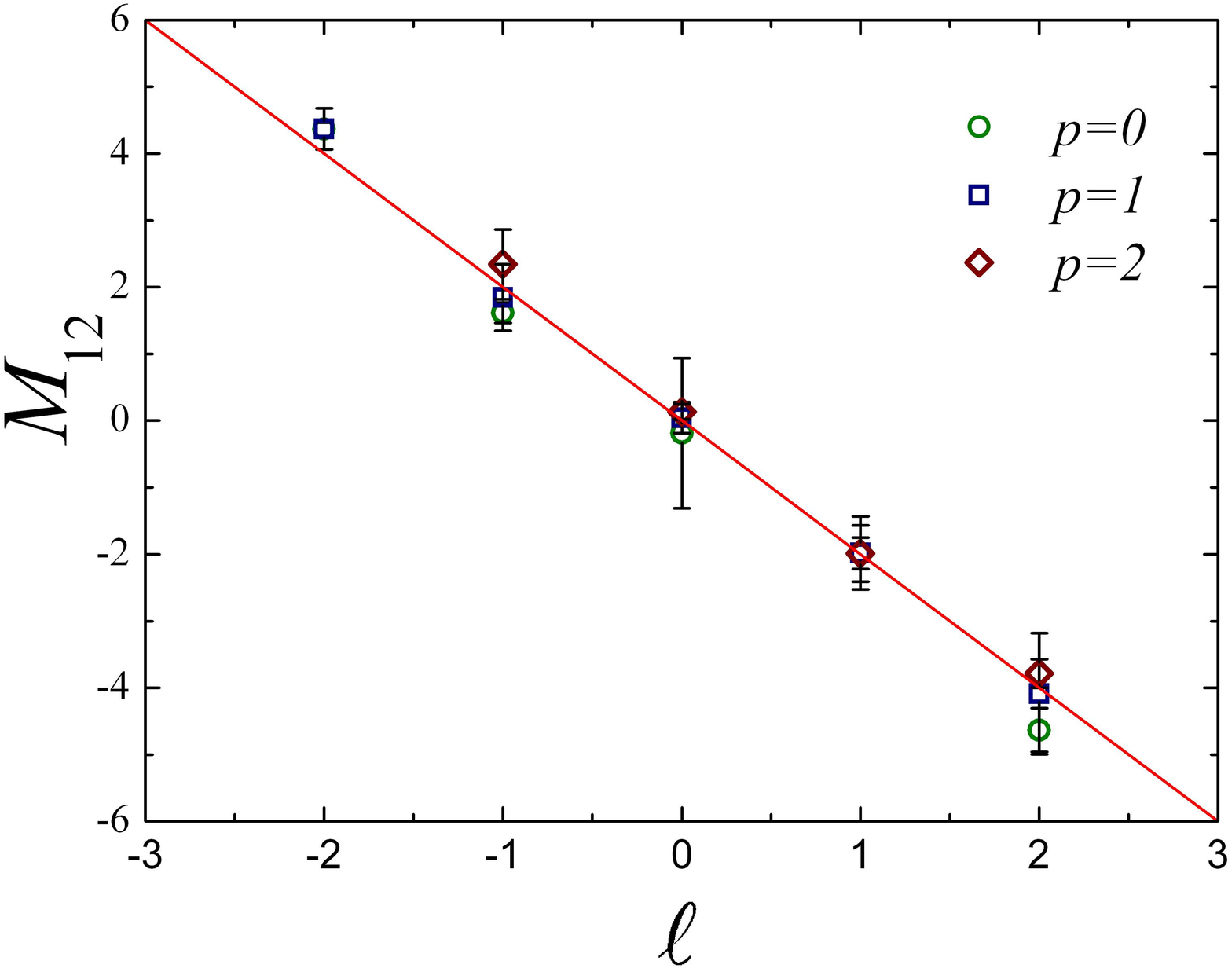}} \caption{(color online) Matrix element for different values of $\ell$ and
$p$. The solid line is $M_{12}=-2\ell$. }
\end{figure}

Figure 2 shows the measured matrix element $M_{12}$ as function of
$\ell$ and with different values of $p$ . The measured values collapse
to the $M_{12}=-2\ell$ line. This is correctly predicted by our theory.
The matrix element changes linearly on $\ell$ but is independent
on $p$. We have experimentally verified that the positional GH shift
is not affected by the radial mode index (not shown).

\begin{figure}[htb]
 \centerline{\includegraphics[width=6.25cm]{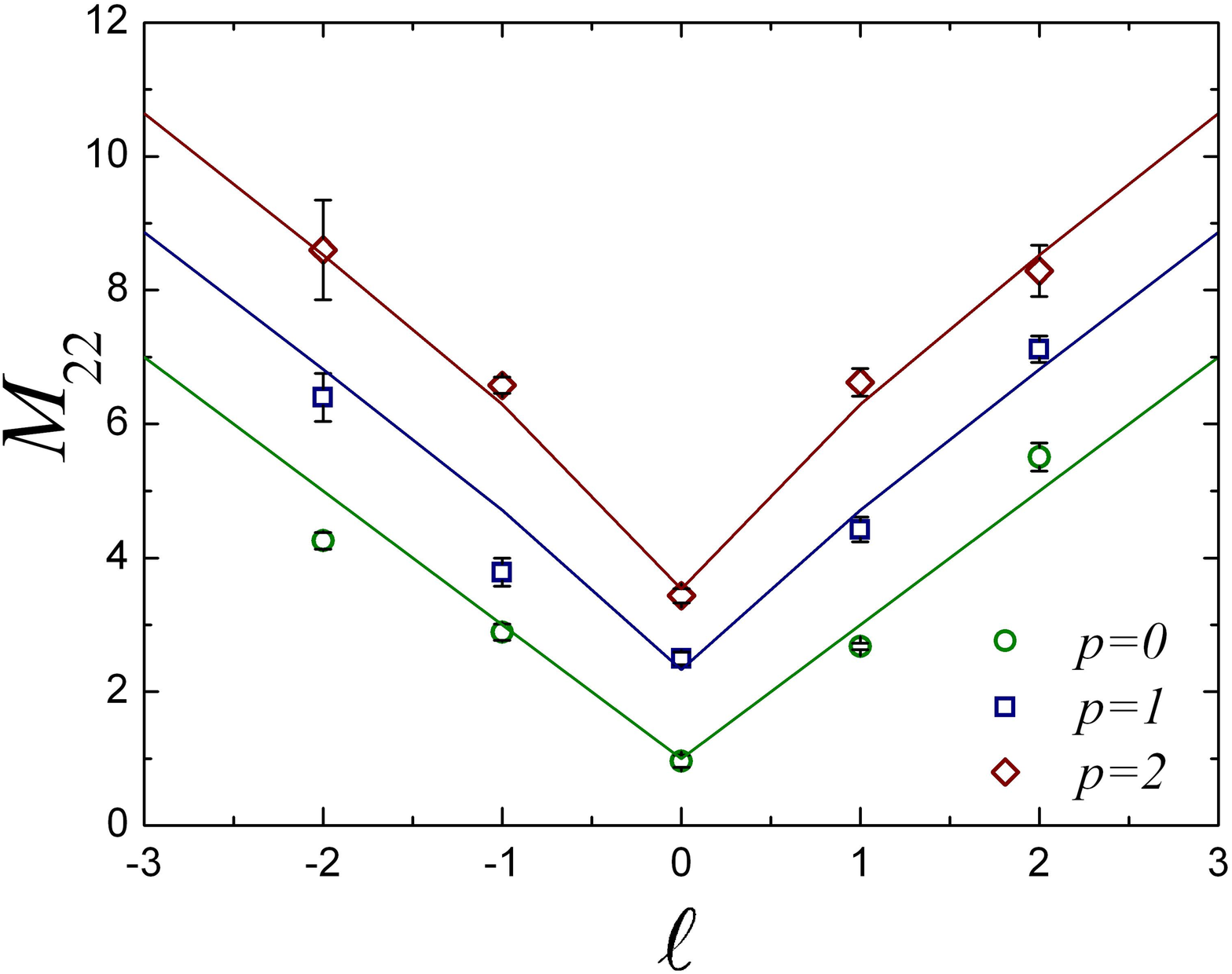}} \caption{(color online) Matrix element $M_{22}$ for different values of $\ell$
and $p$. The solid line is $\xi\left(\ell,p\right)$. A discussion
of $\xi\left(\ell,p\right)$ is in the text.}
\end{figure}

The measured $M_{22}$ is shown in Fig. 3. The theoretical predictions
are calculated from $\xi\left(\ell,p\right)$ for different $\ell$
and $p$ values. For $p>0$ we obtained greater enhancements. This
confirms our theoretical prediction of the effect of $p$ on the angular
shifts. At $p=0$, the angular IF shift linearly increases as $1+2|\ell|$.
Furthermore, had we used a centroid detector the increase of the angular
shift would have been $\xi\left(\ell,p\right)=2p+|\ell|+1$, a linear
increase with $p$.

In conclusion, we have theoretically described and experimentally
measured the positional GH shift and angular IF shifts for higher-order
$LG_{\ell,p}$ to determine the role of mode indices $\ell$ and $p$.
Specifically, we have observed that only the azimuthal index $\ell$
affects the positional shift while the angular shifts are enhanced
by \textit{both} the radial $p$ and the azimuthal $\ell$ indices.
Higher-order $LG_{\ell,p}$ beams form a complete basis set for paraxial
light beams contrary to $LG_{\ell,p=0}$\cite{Siegman}. Hence with
proper mode decomposition, beam shifts associated with arbitrary modes
of light can be predicted based on our theoretical and experimental
results.

This work is supported by the Foundation for Fundamental Research
of Matter (FOM) and the European Union within FET Open-FP7 ICT as
part of STREP Program 255914 Phorbitech.

\pagebreak{}

\section*{Informational Fourth Page}


\begin{thebibliography}{References}
\bibitem{centroid}If a centroid detector is used instead of a median
detector, $2\ell\longrightarrow\ell$ and $\xi\left(\ell,p\right)=2p+|\ell|+1$.
\end{thebibliography}
\begin{thebibliography}{References}
\bibitem{centroid-1}If a centroid detector is used instead of a median
detector, $2\ell\longrightarrow\ell$and $\xi\left(\ell,p\right)=2p+|\ell|+1$
.
\end{thebibliography}
\end{document}